\documentclass[fleqn,usenatbib]{mnras}

\usepackage{newtxtext,newtxmath}

\usepackage[T1]{fontenc}
\usepackage{ae,aecompl}

\usepackage{graphicx}	
\usepackage{amsmath}	
\usepackage{amssymb}	
\usepackage{longtable}    

\newcommand{\Nratio}{$^{14}$N$/$$^{15}$N\,}	

\title[Nitrogen Fractionation in Galaxies]{Nitrogen Fractionation in External Galaxies}

\author[Serena Viti et al.]{
Serena Viti$^{1}$\thanks{E-mail: serena.viti@ucl.ac.uk},
Francesco Fontani$^{2}$,
Izaskun Jim\'enez-Serra$^{3,1}$,
Jonathan Holdship$^{1}$
\\
$^{1}$University College London
Gower Street, London, WC1E 6BT, UK\\
$^{2}$Osservatorio Astrofisico di Arcetri, Largo E. Fermi 2, 
I-50125 Firenze, ITALY\\
$^{3}$Centro de Astrobiolog\'ia (CSIC/INTA),
Ctra de Torrej\'on a Ajalvir, km 4
28850 Torrej\'on de Ardoz, Madrid Spain}

\date{Accepted 2019 April 25. Received 2019 March 25; in original form 2018 December 12}

\pubyear{2015}

\begin{document}
\label{firstpage}
\pagerange{\pageref{firstpage}--\pageref{lastpage}}
\maketitle

\begin{abstract}
In star forming regions in our own Galaxy, the $^{14}$N/$^{15}$N ratio is found to vary from $\sim$ 100 in meteorites, comets and protoplanetary disks up to $\sim$ 1000 in pre-stellar and star forming cores, while in external galaxies the very few single-dish large scale measurements of this ratio lead to values of 100-450.
The extent of the contribution of isotopic fractionation to these variations is, to date, unknown. In this paper we present a theoretical chemical study of nitrogen fractionation in external galaxies in order to determine the physical conditions that may lead to a spread of the $^{14}$N/$^{15}$N ratio from the solar value of $\sim$ 440 and hence evaluate the contribution of chemical reactions in the ISM to nitrogen fractionation. We find that the main cause of ISM enrichment of nitrogen fractionation is high gas densities,  aided by high fluxes of cosmic rays.
\end{abstract}




\section{Introduction}

Nitrogen is the fifth most abundant element in the Universe that can exist in the form of two stable isotopes, $^{14}$N and $^{15}$N. The \Nratio ratio has been measured in Solar-system objects such as comets, meteorites and chondrites (Mumma \& Charnley 2011; F\"uri \& Marty 2015), in molecular clouds with and without the influence of star formation processes (Adande \& Ziurys 2012; Hily-Blant et al. 2013; Bizzocchi et al. 2013; Fontani et al.~2015; Guzm\'an et al. 2017; Zeng et al.~2017; Colzi et al. 2018a, 2018b; Kahane et al. 2018; De Simone et al. 2018; Redaelli et al. 2018), and in galaxies (Henkel et al. 1998, 2018; Chin et al. 1999). In star-forming regions, there is a large spread in the measured \Nratio ratio, ranging from $\sim$100 for meteorites, comets and protoplanetary disks to $\sim$1000 in pre-stellar and star-forming cores. The Solar nebula value measured in the Solar wind and in Jupiter's atmosphere is an intermediate value, around 440 (Fouchet et al. 2004; Marty et al. 2010). In the few extragalactic sources where the \Nratio ratio has  been measured, its values range from $\sim$100-450 (see Table~\ref{obsgalaxies} below). 

The \Nratio ratio is considered a good indicator of stellar nucleosynthesis, since the two isotopes are not synthesized in the same way. Both isotopes are thought to be actively produced in the CNO cycles of massive stars and in the so-called Hot Bottom Burning of asymptotic giant branch (AGB) stars (e.g.~Schmitt \& Ness 2002, Izzard et al. 2004). However, there should be some differences in their nucleosynthesis necessary to explain their observational behaviour, such as the strong primary component of $^{14}$N at low metallicity (e.g. Matteucci~1986), or the relative role played by massive stars and novae in the (over-)production of $^{15}$N with respect to $^{14}$N (e.g. Clayton 2003, Romano \& Matteucci 2003, Prantzos 2011, Romano et al. 2017). The relative importance of these processes, and the existence of additional processes not yet considered, are still unclear. In particular, the contribution of the isotopic fractionation, i.e. the role of chemical reactions occurring in the gas phase of the interstellar medium (ISM; see e.g. Roueff et al. 2015, Wirstr{\"o}m \& Charnley 2018), which are unrelated to stellar nucleosynthesis, has not been explored in detail under the different physical conditions expected in extragalactic environments.
In this work we perform, for the first time, a chemical modelling study of nitrogen fractionation that may be occurring in the gaseous component of external galaxies. In section 2, we present the chemical model and network used for the $^{14}$N and $^{15}$N isotopic species; in Section 3 we present our results for the modelling of the nitrogen fractionation in gas at different H$_2$ densities and extinction, and affected by energetic phenomena (such as stellar heating, UV radiation and cosmic rays). In Section 4, we report our conclusions.

\section{Chemical modelling of Nitrogen fractionation}
\label{model}

The chemical modelling was carried out using the open source time dependent gas-grain chemical code UCLCHEM\footnote{https://uclchem.github.io/}. The code is explained in detail in Holdship et al. (2017). Here we briefly summarize its main characteristics. UCLCHEM computes the evolution, as a function of time, of chemical abundances of the gas and on the ices starting from a diffuse and atomic gas.  We ran UCLCHEM in two phases in a very similar manner as in Viti (2017) where theoretical abundances for extragalactic studies were derived.  In Phase I, the gas is allowed to collapse and to reach a high density by means of a free-fall collapse. The temperature during this phase is kept constant at 10 K, and the cosmic ray ionization rate and radiation field are at their standard Galactic values of $\zeta_o$ = 5$\times$10$^{-17}$ s$^{-1}$ and 1 Draine, or 1.6$\times$10$^{-3}$ erg/s/cm$^2$ (Draine~1978, Draine \& Bertoldi~1996). During Phase I, atoms and molecules are allowed to freeze onto the dust grains and react with each other, forming icy mantles. In Phase II, we compute the chemical evolution of the gas  after some energetic event has occurred (simulating either the presence of an AGN and/or a starburst). 

The initial (solar) elemental abundances considered in our models were taken from Asplund et al. (2009). Our elemental isotopic nitrogen ratio is 440. 
UCLCHEM includes non-thermal desorption processes during the cold phase. While UCLCHEM also includes thermal desorption processes as described in Viti et al (2004), for this work we simply assume instantaneous evaporation for the second phase. 

In both phases, the basic gas phase chemical network is based on the UMIST13 database (McElroy et al. 2013) with updates from the KIDA database (Wakelam et al. 2015). The surface reactions included in this model are assumed to be mainly hydrogenation reactions, allowing chemical saturation when possible. The network contains 2908 reactions and 239 chemical species. The number of reactions is reduced with respect to other networks reproducing the chemistry of molecular cloud/cores (e.g. Loison et al. 2019), but similar to other networks used to reproduce the chemistry of nearby Galaxies (Viti et al. 2014).

%
%

For the $^{15}$N network, we  duplicated the $^{14}$N network changing all $^{14}$N by $^{15}$N. We also added the $^{15}$N exchange reactions used by Roueff et al. (2015, see their Tables 1 and 2), with the only exception of those reactions involving ortho-H$_2$ and para-H$_2$ for which we only used the reaction rate from the ortho-H$_2$ species. This is partially justified because when we calculated the rate for the para and ortho species at 10 and 100~K, we systematically found that the ortho rate was orders of magnitude higher than the para ones. Nevertheless we note that, as some studies show (Hily-Blant et al. 2018; Furuya et al. 2015) in some environments para H$_2$ may be dominant. We have therefore performed a further test where we use the rate for para-H$_2$ instead of the one for the ortho H$_2$, essentially assuming in this way that all the molecular hydrogen is in the para form. The only two reactions affected by this exchange are: \\

$^{14}$N$^+$ + H$_2$ $\rightarrow$ NH$^+$ + H \\

$^{15}$N$^+$ + H$_2$ $\rightarrow$ $^{15}$NH$^+$ + H

which essentially only affect ammonia and the nitrogen hydrides. We discuss this further in Section 3.4.
For the ion-neutral reactions for which Roueff et al. (2015) do not give any reaction rate coefficient, we adopted the standard Langevin value of $10^{-9}$ cm$^{3}$s$^{-1}$ for the forward reaction, as done also by Hily-Blant et al.~(2013). We have not included the reactions considered as improbable in Table~1 of Roueff et al. (2015). Finally, we have also checked and updated (where needed) our network according to the reactions given in Table 3 of Loison et al. (2019).

To test the network, we first ran a model with the same initial conditions as in Wirstr\"om \& Charnley (2018)
They assumed a static core with constant H$_2$ volume density of $10^6$ cm$^{-3}$, constant gas temperature of 10~K, a cosmic ray ionisation rate $\zeta = 3\times 10^{-17}$ s$^{-1}$, and visual extinction $A_{\rm v}=10$. In our test model, we have used these same input parameters, as well as the same initial elemental abundances of C, N and O (taken from Savage \& Sembach~1996). 
Usually, in our model carbon is, initially, totally in the atomic form (C or C$^+$), while Wirstr\"om \& Charnley~(2018) assume it is totally locked in CO. Therefore, we have adapted our model to also reproduce this initial condition. We find values of the HCN abundance w.r.t. H$_2$, and \Nratio in HCN, very similar to those computed by Wirstr\"om \& Charnley~(2018): in our model, the HCN abundance rises up to $\sim 10^{-9}$ at $\sim 10^{4}$ yrs, and then it drops by several orders of magnitude afterwards. We also find that \Nratio for HCN/HC$^{15}$N is about 400, as found by Wirst\"om \& Charnley (2018), and thus conclude that our updated model can reproduce the most recent dark cloud models including $^{15}$N fractionation.
We note that we have not considered the doubly substituted N$_2$, as done by Wirstr\"om \& Charnley (2018), because we have assumed that this species is negligible for the chemistry of extragalactic environments at large scales. Our assumption is likely correct because we are able to reproduce the results of Wirstr\"om \& Charnley (2018), which indicates that the reactions involving the doubly substituted N$_2$ are indeed negligible.

Following the approach of Wirstr\"om \& Charnley (2018), we have replicated all reactions involving $^{14}$N species, including those in which more than one product includes nitrogen. This could lead, at high densities and long times, to an artificial increase of the values of the isotopic ratios. Therefore, we recommend to consider these values at long evolutionary times with caution.

Our initial grid includes 288 models, spanning the following parameter space:  gas final densities from 10$^4$ to 10$^6$ cm$^{-3}$, visual extinctions from 1 to 100 mags, temperatures of 50 and 100~K, radiation fields from 1 to 100 Draine, and cosmic ray ionization rates from 1 to 10$^4$ standard galactic cosmic ray ionization field, all selected to cover the ranges likely to be appropriate for external galaxies. The temperature, radiation field and cosmic ray ionization rate vary only in Phase II. Our parameter space is motivated primarily by considering the parameters that affect most the line intensities of gas tracers in starburst and AGN-dominated galaxies, as predicted by radiative transfer models. We also note that our parameter ranges are consistent with previous studies of the chemistry in external galaxies (e.g. Bayet et al. 2009, 2010).  Note that the cosmic ray ionization flux is also used to `simulate' an enhancement in X-ray flux. As previously noted (Viti et al. 2014), this approximation has its limitations in that the X-ray flux will heat the gas more efficiently than cosmic rays. However, the chemistry arising from these two fluxes should be similar. In addition we have ran a second grid of models, varying the parameter space as above, at a reduced metallicity of half solar, to mimic environments more similar to the Large Magellanic Cloud. While we do not aim at modelling any galaxy in particular, this parameter space ought to cover the range of possible differences between extragalactic environments, where nitrogen fractionation has been measured, and the Milky Way.


\section{Results}
\label{results}

In this Section we describe our model predictions for \Nratio\ by varying crucial physical and chemical parameters of the host Galaxy, and discuss how they compare with observations. We have analysed the following chemical species: HCN, HNC (the only two species for which measurements have been obtained, see Table 1), CN, and N$_2$H$^+$. In the next section, we will start discussing the most abundant species, i.e. HCN and HNC, and its chemically related species CN.





\subsection{Dependence of fractionation to variations in the physical parameters}
\label{dependence}

A summary of the qualitative trends of the \Nratio with time, as a function of  the combination of the physical parameters, is given in Table~\ref{variations}. 
Although we run models for two representative average temperatures of 50 and 100~K, we find that varying the temperature does not lead to significant changes in the model predictions of the \Nratio and hence we shall not discuss the sensitivity to temperature variations further. 

\begin{figure}
{\includegraphics[width=9cm]{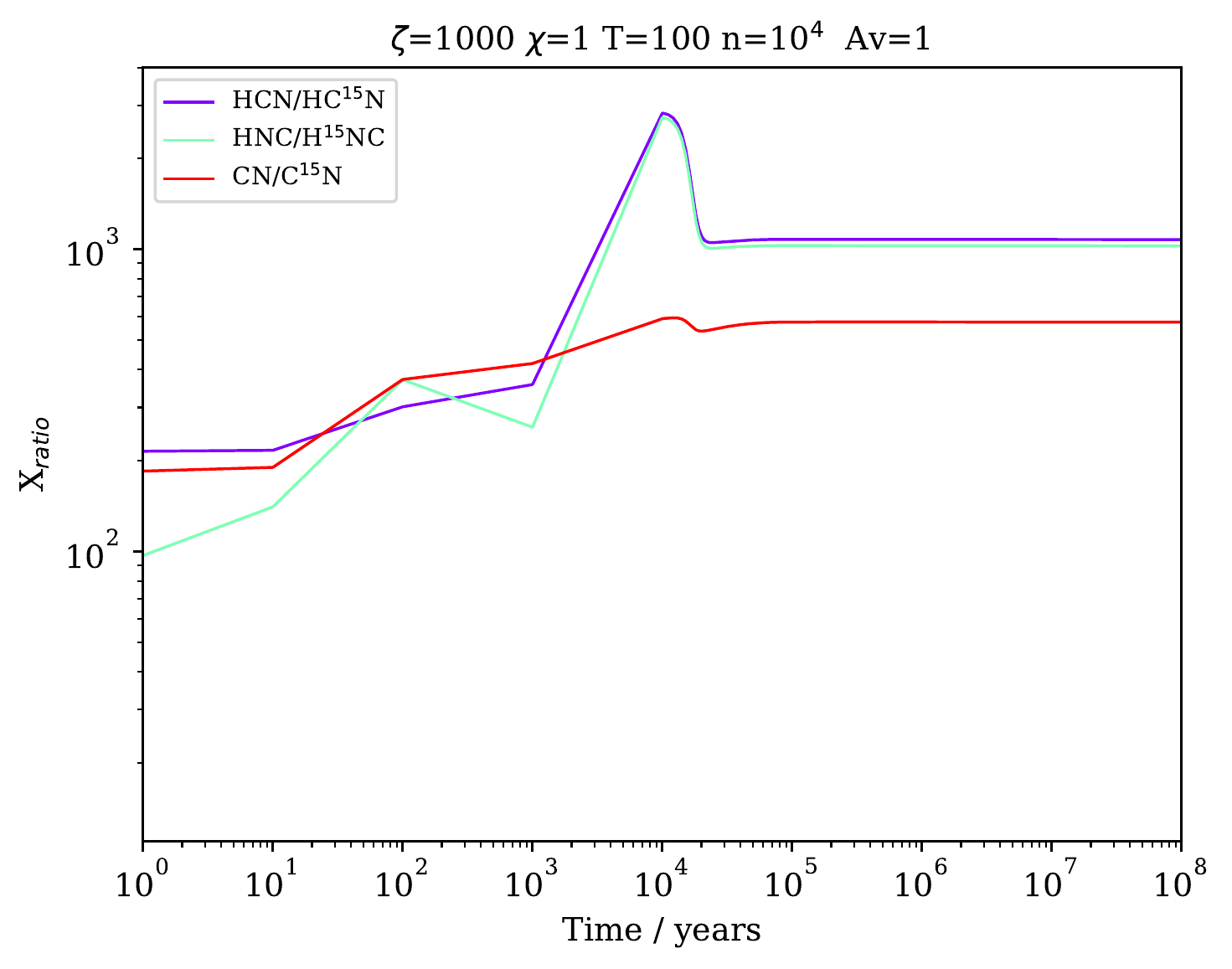}}
{\includegraphics[width=9cm]{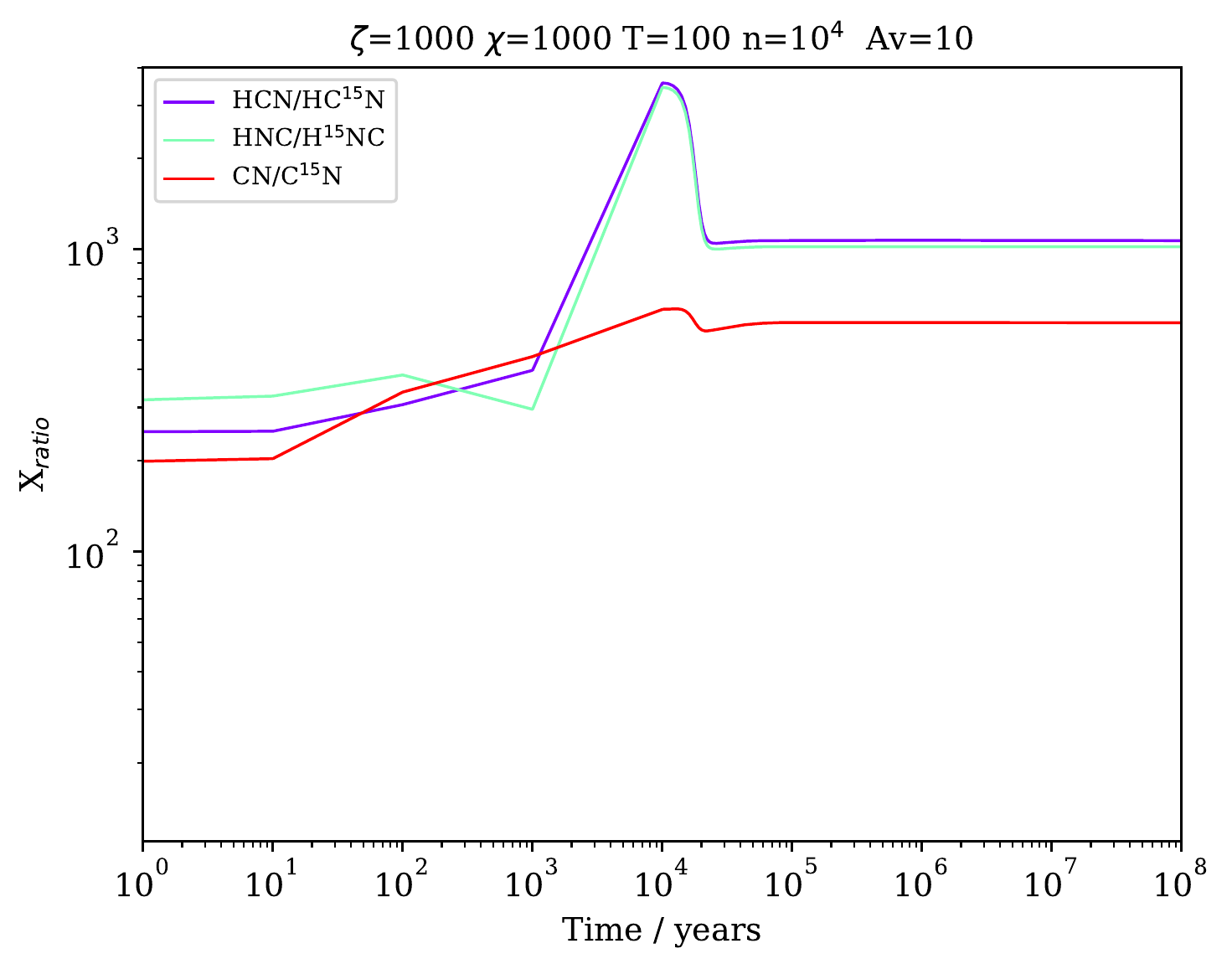}}
     \caption{Plots showing the cases with significant \Nratio increase, i.e. a fractionation decrease. In the title bar $\zeta$ is in units of $\zeta_o$, $\chi$ in units of Draine, the temperature in units of Kelvins, the gas density in units of cm$^{-3}$, and the $A_V$ is in magnitudes. }
     \label{fig:variations_1}
\end{figure}

\begin{figure}
{\includegraphics[width=9cm]{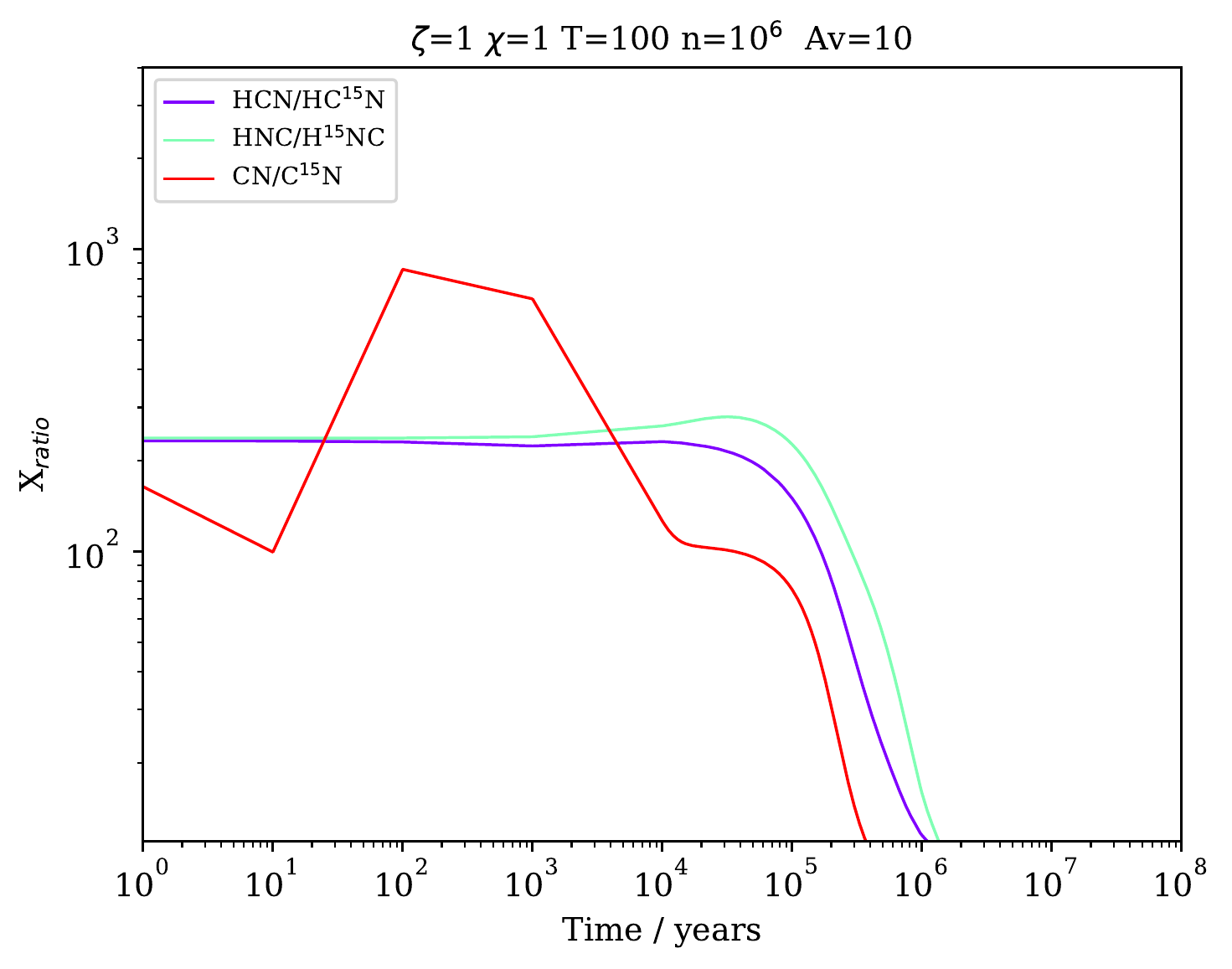}}
{\includegraphics[width=9cm]{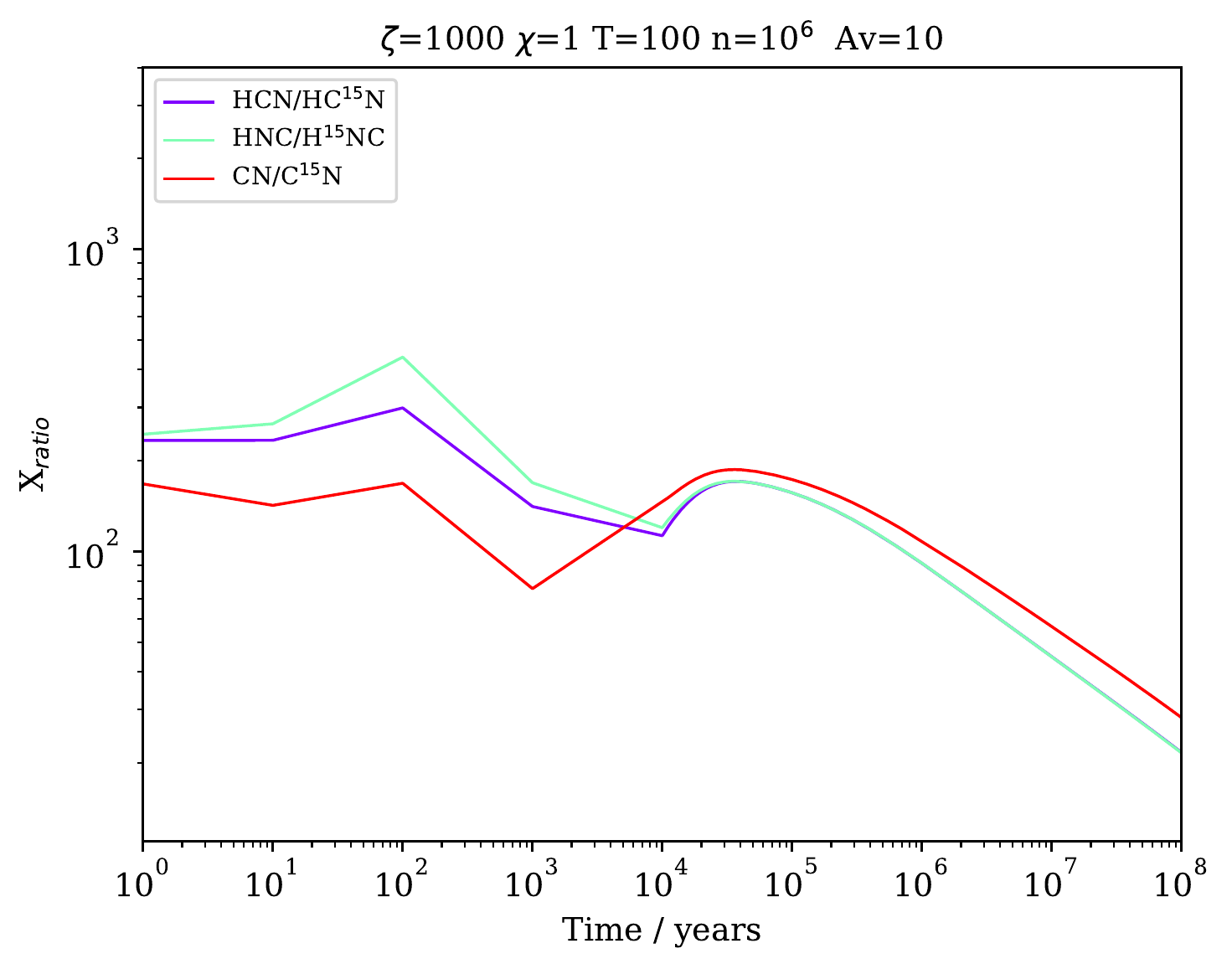}}
{\includegraphics[width=9cm]{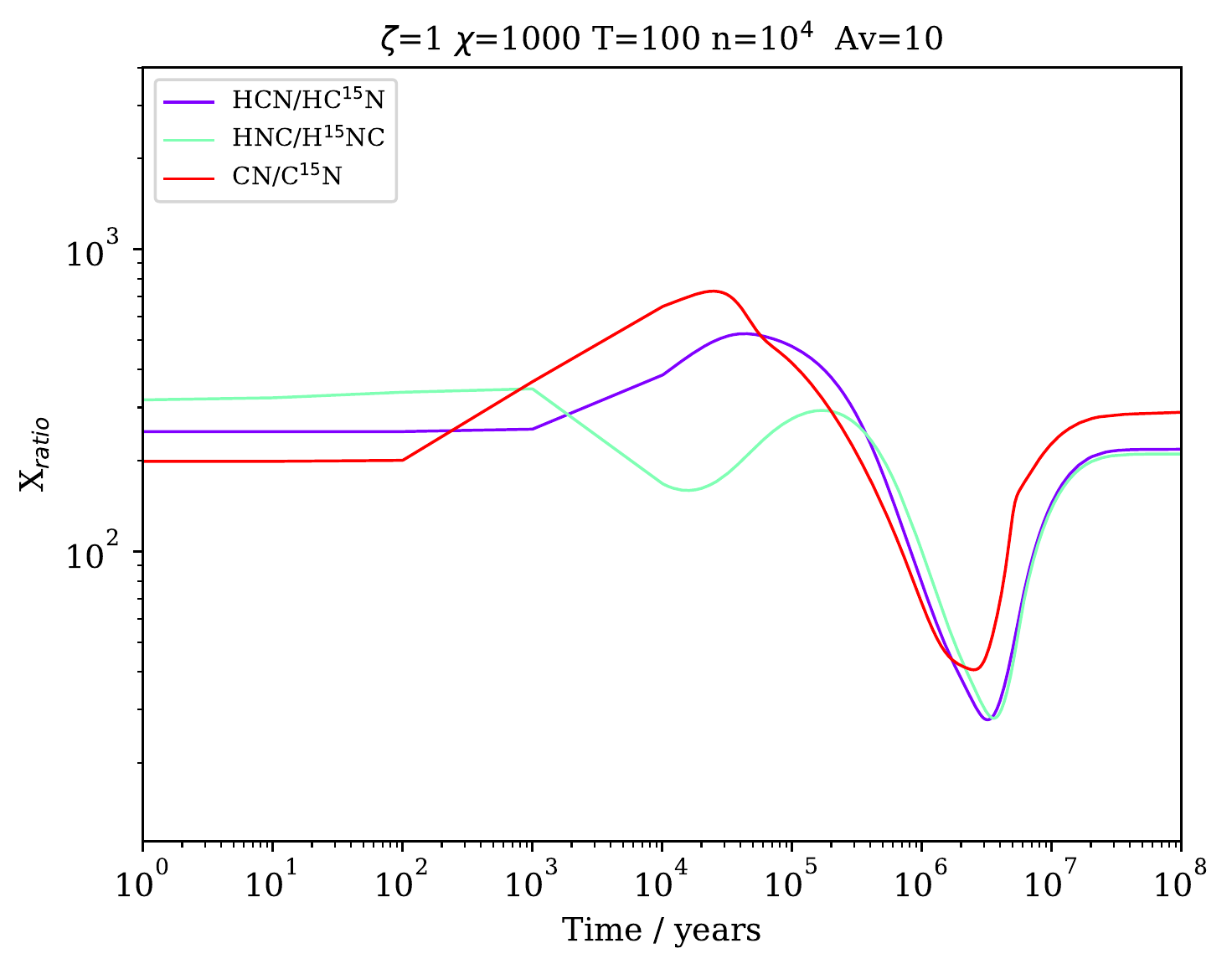}}
     \caption{Plots showing the cases with significant \Nratio decrease, i.e. a fractionation increase. Units as in Figure 1.}
     \label{fig:variations_2}
\end{figure}

\begin{figure}
{\includegraphics[width=9cm]{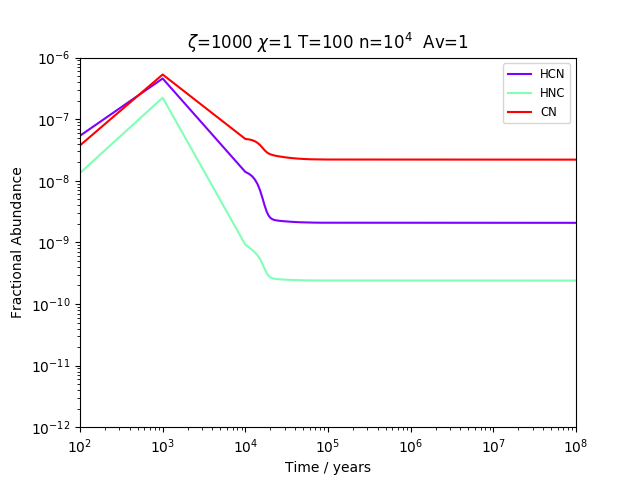}}
{\includegraphics[width=9cm]{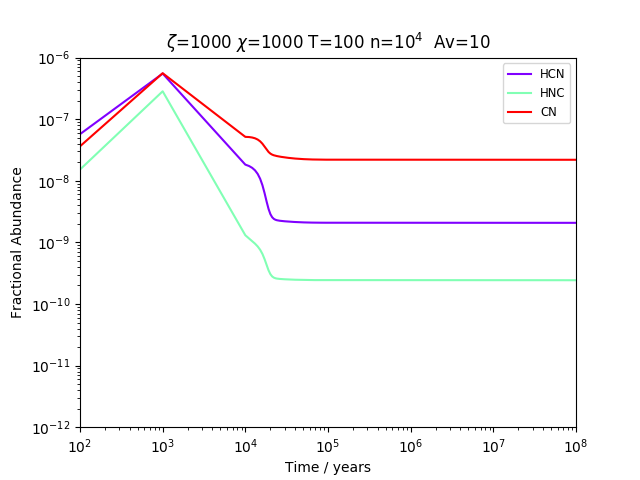}}
     \caption{Plots showing the fractional abundances with respect to the total number of hydrogen nuclei of the main isotopologues of the models of Fig.~\ref{fig:variations_1}.}
     \label{fig:frac1}
\end{figure}

\begin{figure}
{\includegraphics[width=9cm]{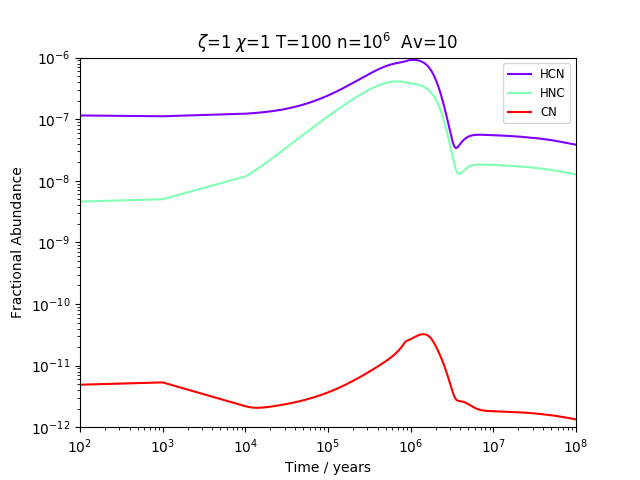}}
{\includegraphics[width=9cm]{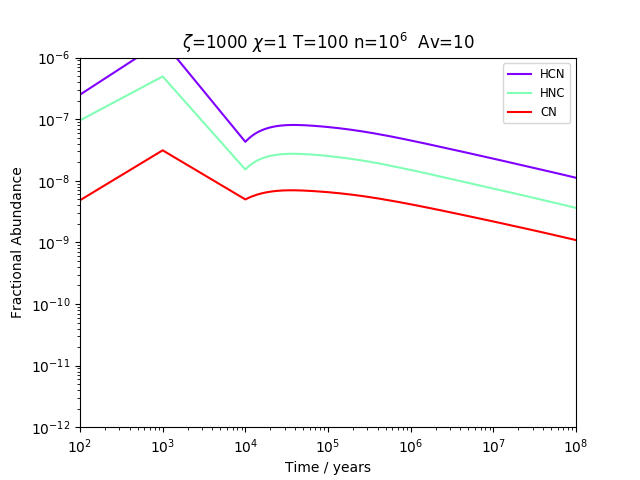}}
{\includegraphics[width=9cm]{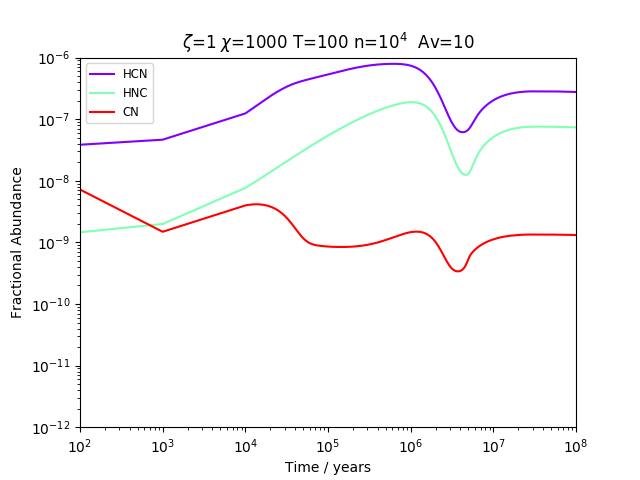}}
     \caption{Plots showing the fractional abundances of the main isotopologues with respect to the total number of hydrogen nuclei of the models of Fig.~\ref{fig:variations_2}.}
     \label{fig:frac2}
\end{figure}

Depending on the combinations of the various parameters, the largest variation with time that we find in \Nratio for HCN, HNC or CN is of an order of magnitude in a range from $\sim 10$ to $\sim 1000$.
In Figs.~\ref{fig:variations_1} and \ref{fig:variations_2} we plot the predictions for \Nratio\ against time showing the largest \Nratio\ increase or decrease, respectively, while in Fig.~\ref{fig:frac1} and Fig.~\ref{fig:frac2} we show the fractional abundances (with respect to the total number of hydrogen nuclei) of the main isotopologues for the same models. Fig.~\ref{fig:variations_1} shows that the largest increase is obtained either when $\zeta=1000$ or when both $\zeta$ and $\chi$ are about 1000 times their standard values, and $A_{\rm V}\geq10$ mags. In all cases, the average density is low ($10^4$ cm$^{-3}$). This means that, at large Giant Molecular Cloud scales (i.e for $n_H$ $\sim$ 10$^4$ cm$^{-3}$), in galaxies with sources of energetic particles such as AGNs or ULIRGs the fractionation should be suppressed with time. 
On the other hand, the highest drop in \Nratio\  (Fig.~\ref{fig:variations_2}) is found for two cases: if $\chi$ is low (1 Draine) but the gas density is high ($10^6$ cm$^{-3}$, top panel in Fig.~\ref{fig:variations_2}), or when $\chi$ and $A_v$ are high (1000 Draine and $\geq$ 10 mags respectively) and the density is low ($10^4$ cm$^{-3}$, bottom panel in Fig.~\ref{fig:variations_2}). A smaller but significant decrease is obtained also when $\zeta$ is high (1000) at high density (middle panel Fig.~\ref{fig:variations_2}). 

We note that in the top and middle panels of Fig. 2, the ratios do not seem to reach a steady state but show a gradual decrease. We have quantified this decrease and found that in reality this is less than 5$\%$, and  likely due to the precision of our calculations. The decrease of the ratios at long times appears large only because the logarithmic Y-scale tends to magnify the changes that occur at low ratios.  

The above discussion describes our analysis of solar metallicity models. 
As mentioned in Section 2, we also ran models for metallicities half the solar one in order to reproduce the possible trend in a Galaxy like the Small Magellanic Cloud, or other low metallicity galaxies. In general, we do not find any significant difference in the trends. For some of the models we find slightly different absolute values of the \Nratio but the range remains the same.


\subsection{Differences in fractionation among N-bearing molecules}
\label{diff_molecules}

One of the clearest results from our modelling is that HCN and HNC show little differences in their \Nratio within a factor 2. The fractionation of CN, on the other hand, shows more variability, especially with time for many models. 
In particular, for cosmic ray ionization rates $\geq$ 1000 the standard one, and densities $\leq$ 10$^5$ cm$^{-3}$, the  CN fractionation at late times is always higher than that of HNC and HNC by more than a factor of 2. 



\subsection{Comparison with observations}
\label{observations}

In Table~\ref{obsgalaxies}, we list the observational values of \Nratio\ for all external galaxies reported in the literature. 
As reference, in Table~\ref{obsmilkyway} we also list the average values (with the dispersions) of the \Nratio\ obtained in massive star-forming clumps and diffuse clouds in the Milky Way. The Milky Way can be considered as a template for spiral Galaxies, hence these clumps represent a proxy of the densest portions in spirals. We do not include in the table low-mass star-forming cores.

Unfortunately, the only two species detected in the $^{15}$N isotope in external galaxies are HCN and (in fewer places) HNC. Hence, we focus the comparison with our models  on HCN. Our criterion for choosing the models that best reproduce the observations is that the ratio has to be matched by 10$^5$ years and be maintained up to a million year. For the galaxies where we only have a lower limit for this ratio, we have imposed an arbitrary upper limit of 1000. 

We note that in general many models match the observed value of fractionation, indicating that the observed ratio is achievable under a large range of physical and chemical parameters. More specifically, for both NGC4945 and Arp220, the range of observed values is achieved by models of gas at low visual extinction for  gas densities up to 10$^5$ cm$^{-3}$ and cosmic ray ionization rate up to 100 the standard value. However for NGC4945 there are also some models at  high densities (10$^6$ cm$^{-3}$) at all cosmic ray ionization rates that can match the observed ratio at low and high visual extinctions. For Arp220, densities of 10$^6$ cm$^{-3}$ can only match observations for the highest cosmic ray ionziation rates and highest radiation fields at low visual extinctions. This may in fact be consistent with the high star formation rates found in the nuclear region of this galaxy.
We note that only for these high densities the radiation field has an impact on the fractionation ratio. IC694 has similar ranges of fractionation to NGC4945 but with a lower upper limit and this indeed reduces the best matches among the models: here only densities $\geq$ 10$^5$ cm$^{-3}$ fit the observed fractionation and, for 10$^5$ cm$^{-3}$, only at visual extinctions of 1 mag for cosmic ray ionization rates and radiation fields of up to 100 and 10 the standard value, respectively. For higher densities, higher values of radiation, cosmic rays and, in some cases, visual extinction, also match the ratio. For the galaxies where we only have lower limits, even imposing an upper limit of 1000, leads to too many models matching the ratio to discuss them here. For the LMC, on the other hand, we are able to constrain the physical parameters much better, as there are only very few models that match the observations: a model with  a gas density of 10$^5$ cm$^{-3}$ with a standard galactic cosmic ray ionization rate and an Av of $>$ 10  mags,  or models with a density of 10$^6$ cm$^{-3}$, Av$\geq$ 10 mags and $\zeta$ $>$ 100 the standard value. In fact, the measured extinction in the LMC is significantly lower than the average found in the Milky Way (Dobashi et al. 2008), so the first case may be favoured. The radiation field is not constrained. 

Together with the results from Sections 3.1 and 3.2, we can conclude that the main cause of enrichment in $^{15}$N is high densities, but it can be aided by high fluxes of cosmic rays and, to a lesser extent, an intense radiation field.


\begin{table*}
\caption{\Nratio measured in external galaxies}
\begin{tabular}{|ccccc|}
\hline \hline
Galaxy  & type & $^{14}$N/$^{15}$N  &  Molecule & Reference  \\           
\hline \hline
NGC4945 & starburst & 200-500 & HCN & Henkel et al. (2018) \\ 
LMC & 0.5 metal & 111($\pm$ 17) & HCN & Chin et al. (1999) \\
Arp220 & ULIRG & 440 (+140,-82) & HCN, HNC & Wang et al. (2016) \\
NGC1068 & AGN+starburst & $>$ 419 & HCN & Wang et al. (2014) \\
IC694 & starburst & 200-400(?) & HCN & Jiang et al. (2011) \\
LMC & 0.5 metal & 91($\pm$ 21) & HCN & Wang et al. (2009) \\
M82 & starburst & $>$ 100 & HCN & Henkel et al. (1998) \\
Galactic Center & standard with high $\zeta$ & $\geq$164 & HNC & Adande \& Ziurys (2012) \\ \hline
\end{tabular}
\label{obsgalaxies}
\end{table*}



\subsection{Fractionation predictions for N$_2$H$^+$ and NH$_3$ in external galaxies}

Not many nitrogen bearing species have been observed to be abundant in external galaxies. Beside HCN, HNC and CN, discussed already in previous sections, the most common nitrogen bearing species detected in nearby galaxies are: HNCO, HC$_3$N, CH$_3$CN, and N$_2$H$^+$. While our network does include all the $^{14}$N isotopologues of these species, a fractionation chemistry for the first three of these species is not available, and hence we concentrate on the predicted fractionation of N$_2$H$^+$, an important tracer of cold and dense gas. 

Aladro et al. (2015) detected N$_2$H$^+$ in 4 galaxies: M83, NGC253, M82, and M51, and found a column density of 6.5$\times$10$^{12}$, 4$\times$10$^{13}$, 1$\times$10$^{13}$ and 4$\times$10$^{12}$ cm$^{-2}$, respectively. Grouping M83 with M51 and M82 with NGC253 (due to their similar values of observed N$_2$H$^+$) we find that for the first two galaxies this translates into a N$_2$H$^+$ fractional  abundance  ranging between 2.5 and 4$\times$10$^{-10}$ if the visual extinction is 10 mag and 2.5 and 4$\times$10$^{-11}$ if the visual extinction is 100 mag. For the other two galaxies, we get an abundance of $\sim$6.2$\times$10$^{-10}$ to 2.5$\times$10$^{-9}$ for 10 mag and $\sim$6$\times$10$^{-11}$ to 2.5$\times$10$^{-10}$ for 100 mag. In order to predict the expected fractionation of N$_2$H$^+$ in these galaxies we restrict our grid of models to those that match these abundances. 

{\it M83 and M51}: we find that, if the visual extinction traced by N$_2$H$^+$ is close to 10 mags, then two  models can reproduce the range of abundances but both only for {\it a short} period of time, in some cases as brief as 1000 years: a model with a cosmic ray ionization rate higher than the galactic standard one by a factor of 1000, a gas temperature of 50 K and a gas density of 10$^4$ cm$^{-3}$, and another model with a cosmic ray ionization rate higher than the galactic standard one by a factor of 10000, a gas temperature of 100 K and a gas density of 10$^5$ cm$^{-3}$.  Clearly, N$_2$H$^+$ is tracing dense gas but it is interesting to note that only high levels of cosmic ray ionization rate can maintain its abundance if the temperature of the gas is $>$ 10 K. 
If the gas has a visual extinction of 100 mags, then the only models that achieve to maintain a high abundance of N$_2$H$^+$ have a cosmic ray ionization rate of 100 times that of the galactic one, a temperature of 50 or 100 K and a gas density of 10$^4$ cm$^{-3}$. In this case, however, N$_2$H$^+$ is not destroyed before 10,000 years. We note that both M51 and M83 are spiral galaxies, with M83 being a young starburst and M51 having recently interacted with a nearby galaxy triggering star formation. Hence both are likely to have an enhanced cosmic ray ionization rate. 

{\it M82 and NGC253}: at 10 mags the only model that reproduces the observed abundance of N$_2$H$^+$ is one with a high cosmic ray ionization rate (1000 $\zeta_o$), a temperature of 100 K and a gas density of 10$^4$ cm$^{-3}$, while if the gas is at a visual extinction of 100 mags then the same model but with a factor of 10 less cosmic ray ionization rate can reproduce the observed abundance of N$_2$H$^+$. We recall that the derived abundances from the observations are different at different extinctions which is why for this comparison models at different visual extinctions do give different matches. We note that these two galaxies are the prototypical chemically rich starburst galaxies and, again, as for the other two galaxies, a higher than standard cosmic ray ionization rate is expected. We also note that while the abundance of N$_2$H$^+$ is not sensitive to changes in the radiation field, for all our best fit models the latter can not exceed $\sim$ 100-1000 Draine. 

These results indicate that  either  the observed N$_2$H$^+$ is tracing in fact colder gas than we modelled, or that it is indeed tracing gas close to a source of high cosmic ray flux that maintains its abundance for longer. In order to exclude the former hypothesis we ran a test model whereby we maintained in Phase 2 all the parameters as in Phase 1 (including the temperature of the gas at 10 K) and ran the model for 10$^7$ years. We find that, regardless of the gas density, we can not obtain an N$_2$H$^+$ abundance much higher than 10$^{-11}$ (which is below the observational value for most observations) for times less than 1 million years (see Figure 5). Hence we conclude that the high abundance of N$_2$H$^+$ is indeed a consequence of the cosmic ray ionization rate but that it is indeed transient implying that the gas traced by N$_2$H$^+$ and observed by Aladro et al. (2015) is  young or most likely replenished periodically. Our predictions also imply that high N$_2$H$^+$ abundances are preferentially seen towards young galaxies, and thus N$_2$H$^+$ could be potentially an evolutionary indicator, although this conclusion has to be taken with caution given the large number of parameters that should produce the predicted high N$_2$H$^+$ abundance.

\begin{figure}
\hspace{-1.6cm}
{\includegraphics[width=7cm, angle=-90]{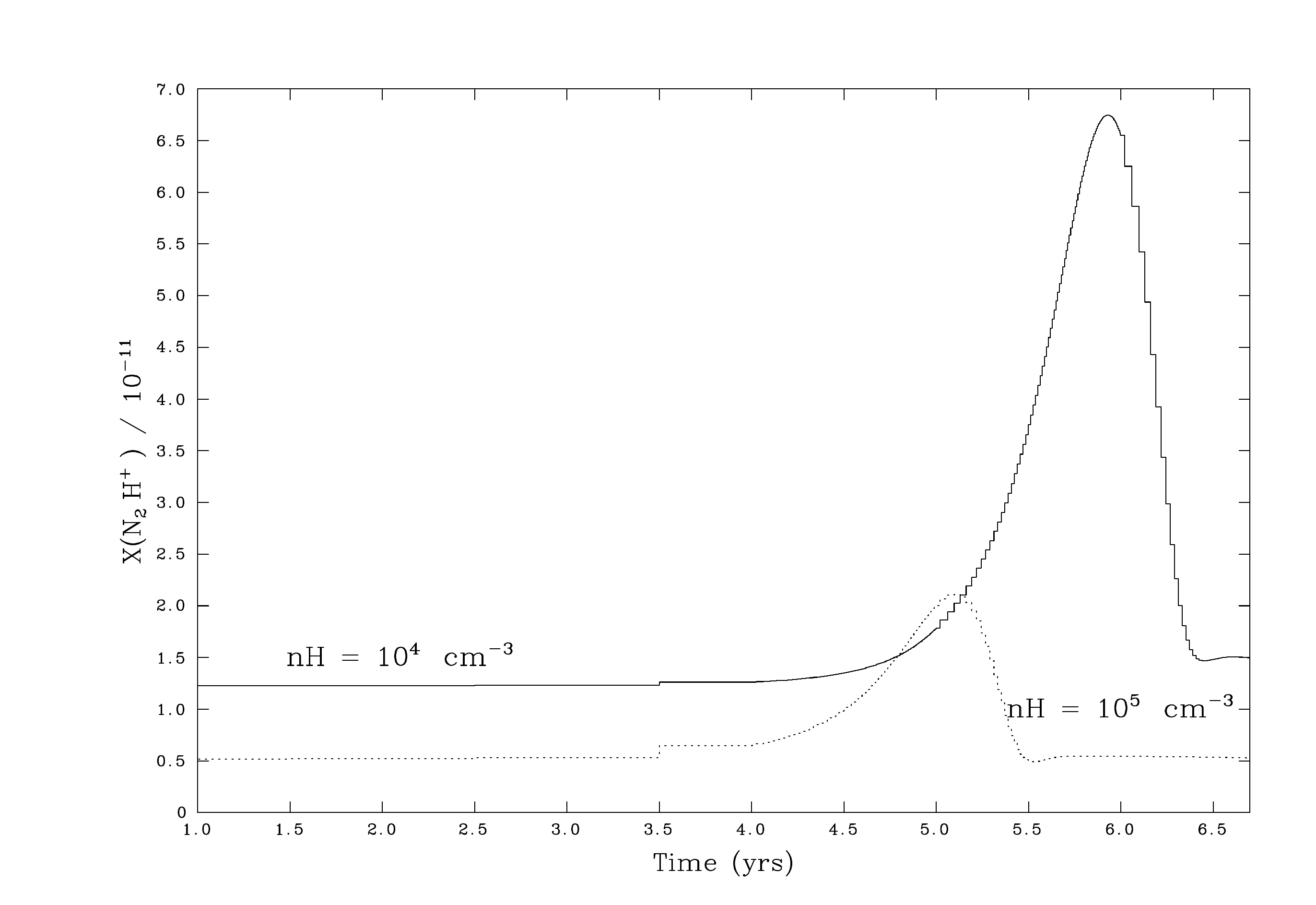}}
\caption{Fractional abundance of N$_2$H$^+$ as a function of time for Phase 2 of two models varying in gas densities, at a constant temperature of 10K (see text).}
     \label{fig:n2hp}
\end{figure}

What do these best matching models predict in terms of fractionation? Surprisingly the ratio of N$_2$H$^+$ to either of its fractionated counterparts is {\it always} at least 10$^4$ implying an extremely low fractionation. 
Assuming that our chemical network for the fractionation of N$_2$H$^+$ is complete, it is therefore  unlikely we would be able to detect $^{15}$NNH$^{+}$ or N$^{15}$NH$^+$ in a reasonable
amount of integration time even with the current, most powerful
facilities.

\begin{figure}

{\includegraphics[width=9cm, angle=0]{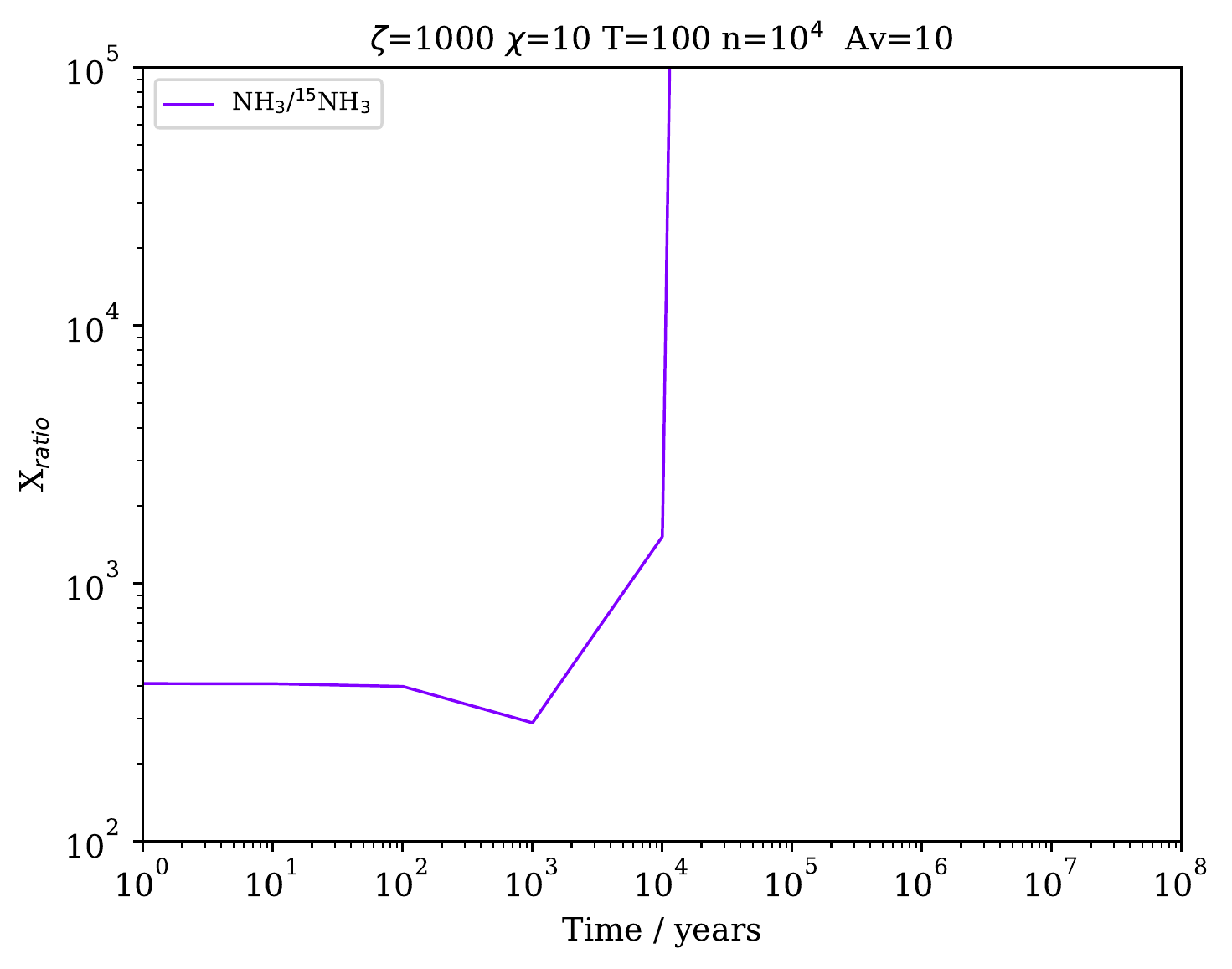}}
\caption{NH$_3$ fractionation for one of the best fit model for NGC~253 as derived from the N$_2$H$^+$ observations. }
     \label{fig:nh3}
\end{figure}

Finally it is worth briefly discussing our predictions for NH$_3$ fractionation. Ammonia was detected in NGC~253 (Ott et al. 2005), yielding a
temperature $\leq$ 17-85 K. We plot in Fig.~\ref{fig:nh3} the ammonia isotopic ratio expected in the best fit model for NGC~253. As we can see from the figure, for gas older than 10,000 years the predicted fractionation is far too low to be detectable.  As mentioned in Section 2, omitting the reactions with para-H$_2$ may have consequences for the abundance of hydrogen nitrides. We therefore  compared the NH$_3$ gas fractional abundance between our model and the one  performed assuming all the molecular hydrogen in the para form, as described in Section 2, but found that while they differ by a factor of two at the end of the cold phase, they are essentially the same in Phase 2. This is  because once the gas is heated to $\sim$ 100 K, the ammonia formed on the ices via hydrogenation is released back to the gas phase and any difference in its abundance between the two models disappears.

\label{predictions}


%
%
%
%
%

%
\begin{table*}
	\centering
	\caption{Qualitative trends of fractionation  as a function of different parameters.}
	\label{tab:comparison}
	\begin{tabular}{lcccc} 
		\hline
		Model & \multicolumn{2}{c}{$A_{\rm V}=1$ mag} &  \multicolumn{2}{c}{$A_{\rm V}\geq 10$ mag} \\
		\hline
		standard & \multicolumn{2}{c}{Constant, apart from a transient enrichment}  &  \multicolumn{2}{c}{\Nratio decrease of 1 order of mag after $10^6$yrs} \\
         & \multicolumn{2}{c}{more pronounced for HCN}  &  \multicolumn{2}{c}{especially at high density} \\
        high $\zeta$ & \multicolumn{2}{c}{Decrease of fractionation  at low densities with time,} & \multicolumn{2}{c}{Decrease/increase of fractionation} \\
         & \multicolumn{2}{c}{flat at high density} & \multicolumn{2}{c}{at low/high density (respectively) with time} \\
		high $\chi$  &  \multicolumn{2}{c}{Constant with time at both densities}  & \multicolumn{2}{c}{Fractionation increase with time for both densities} \\
        high $\zeta + \chi$ & \multicolumn{2}{c}{Constant with time at both densities} & \multicolumn{2}{c}{Fractionation decrease/increase} \\
         & & & \multicolumn{2}{c}{at low/high density, respectively} \\
		\hline
	\end{tabular}
    \label{variations}
\end{table*}
\begin{table*}
\caption{\Nratio measured in the Milky Way in dense and diffuse clouds from different molecules.}
\begin{tabular}{ccccc}
\hline \hline
Reference & \multicolumn{4}{c}{$^{14}$N/$^{15}$N} \\
          & HCN & HNC & CN & N$_2$H$^+$ \\
\hline
Adande \& Ziurys (2012) & & $\sim 130 - 400$ & $\sim 120 - 380$ & \\
Fontani et al. (2015) & & & 190 -- 450 & 180 -- 1300 \\
Ritchey et al. (2015) & & & 274$\pm 18$ & \\
Colzi et al. (2018b) & 115 -- 1305 & 185 -- 780 & & \\
Zeng et al. (2018)  & 70 -- 763 & 161 -- 541 & & \\
\hline
\end{tabular}
\label{obsmilkyway}
\end{table*}
\section{Conclusions}
We have used a time dependent gas-grain chemical model to determine the nitrogen fractionation of dense gas under a range of physical parameters representing galaxies with intense FUV or cosmic ray sources. We determine the sensitivity of the fractionation to the local physical conditions, as well as the fractionation differences  among the observable nitrogen bearing species; we qualitatively test our models by comparing our findings with the few observations of HCN  available and we then make some predictions related to the fractionation for an important nitrogen-bearing species, N$_2$H$^+$. We summarize our findings below:
\begin{itemize}
\item In general we find that in most models the \Nratio for HCN, HNC or CN never varies by more than an order of magnitude with time, and remains in a range from $\sim 100$ to $\sim 1000$.
\item An increase in fractionation can occur at low radiation fields and high densities and viceversa, as well as when both the cosmic ray ionization rate and the gas density are high.
\item A decrease in fractionation is obtained at low densities, high visual extinction and high fluxes of either radiation fields or cosmic rays. 
\item HCN and HNC show little differences in their \Nratio within a factor of 2. On the other hand the \Nratio for CN can be different from that of the other two species at late times for densities $\leq$ 10$^5$ cm$^{-3}$ and cosmic ray ionization rates $\geq$ to 1000 the standard one. 
\item Our models succeed in reproducing the observed  \Nratio in external galaxies but due to the large ranges observed we are unable to fully constrain the physical parameters of  each galaxy with the exception of the LMC whose nitrogen fractionation implies a gas density of 10$^5$ cm$^{-3}$ with galactic cosmic ray ionization rate and an Av of 100 mags, or  a density of 10$^6$ cm$^{-3}$, Av $>$ 10 mags and $\zeta$ $>$ 100. 
\item Finally we predict that even with the most sensitive instruments to date it is unlikely that  we would be able to detect $^{15}$NNH$^{+}$ or N$^{15}$NH$^+$ in external galaxies as their fractionation is more than one order of magnitude lower than that for HCN, HNC or CN.

\end{itemize}

\section*{Acknowledgements}

SV and JH acknowledge STFC grant ST/M001334/1. I.J.-S. acknowledges partial support by the MINECO and FEDER funding under grants ESP2015-65597-C4-1 and ESP2017-86582-C4-1-R.
We are grateful to E. Wirstr\"om and J.-C. Loison for providing us useful clarifications about their models, to C. Henkel for his critical reading of the manuscript, and to the anonymous referee for constructive comments that improved the manuscript.









\bsp	
\label{lastpage}
\end{document}